\documentclass[article]{aa}

\usepackage{natbib}
\usepackage{graphicx}
\usepackage{txfonts}

\begin{document}

\title{Selective spatial damping of propagating kink waves \\ due to resonant absorption}

\author{Terradas$^{1}$, J., Goossens$^{2}$, M., Verth$^{2}$, G.}

\offprints{J. Terradas, \email{jaume.terradas@uib.es}}
\institute{$^1$Departament de F\'\i sica,
Universitat de les Illes Balears, Spain, \email{jaume.terradas@uib.es}
\\$^2$Centre Plasma Astrophysics and Leuven Mathematical Modeling and
Computational Science Centre, Katholieke Universiteit Leuven, Leuven, B-3001,
Belgium, \email{Marcel.Goossens@wis.kuleuven.be, Gary.Verth@wis.kuleuven.be}}
{}

\date{Received / Accepted }

\abstract{There is observational evidence of propagating kink waves driven by
photospheric motions. These disturbances, interpreted as kink magnetohydrodynamic
(MHD) waves are attenuated as they propagate upwards in the solar corona.}{To show
that resonant absorption provides a simple explanation to the spatial damping of
these waves.}{Kink MHD waves are studied using a cylindrical model of solar
magnetic flux tubes which includes a non-uniform layer at the tube boundary.
Assuming that the frequency is real and the longitudinal wavenumber complex, the
damping length and damping per wavelength produced by resonant absorption are
analytically calculated in the thin tube (TT) approximation, valid for coronal
waves. This assumption is relaxed in the case of chromospheric tube waves and
filament thread waves.}{The damping length of propagating kink waves due resonant
absorption is a monotonically decreasing function of frequency.  For kink waves
with low frequencies the damping length is exactly inversely proportional to
frequency and we denote this as the TGV relation. When moving to high frequencies
the TGV relation continues to be an exceptionally good approximation of the actual
dependency of the damping length on frequency. This dependency means that resonant
absorption is selective as it favours low frequency waves and can efficiently
remove high frequency waves from a broad band spectrum of kink waves. The
efficiency of the damping due to resonant absorption depends on the properties of
the equilibrium model, in particular on the width of the non-uniform layer and the
steepness of the variation of the local Alfv\'{e}n speed.}{Resonant absorption is
an effective mechanism for the spatial damping of propagating kink waves. It is
selective as the damping length is inversely proportional to frequency so that the
damping becomes more severe with increasing frequency. This means that radial
inhomogeneity can cause solar waveguides to be a natural low-pass filter for
broadband disturbances. Hence kink wave trains travelling along, e.g., coronal
loops, will have a greater proportion of the high frequency components dissipated
lower down in the atmosphere. This could have important consequences with respect
to the spatial distribution of wave heating in the solar atmosphere.}

\keywords{Magnetohydrodynamics (MHD) --- Waves --- Magnetic fields --- Sun:
atmosphere---Sun: oscillations}

\titlerunning{Spatial damping of propagating kink waves}
\authorrunning{Terradas et al.}
\maketitle

\section{Introduction}

The first observations of post-flare transversal coronal loop oscillations by the
Transition Region and Coronal Explorer (TRACE)
\citep[e.g.,][]{aschetal99,nakaetal99,aschetal02}, inspired much development in
magnetohydrodynamic (MHD) wave theory. This observational breakthrough was
important since estimated wave parameters, such as frequency and amplitude allowed
us to implement magnetoseismological techniques to probe the plasma fine structure
of the Sun's atmosphere, an idea initially proposed by e.g., \citet{uchida1970} and
\citet{robetal84}. It is now commonly accepted that these transversal waves are the
kink mode from MHD wave theory \citep[see e.g.,][]{edrob83}, a highly magnetically
dominated, i.e., Alfv\'{e}nic wave \citep[see][for a discussion on the nature of
kink waves]{goossensetal09}. The observed post-flare kink waves in coronal loops
have two main defining characteristics, firstly they are standing modes and
secondly, they are strongly damped oscillations \citep[in about 1-4 periods, see
e.g.,][]{aschetal03}. Initially there were several physical mechanisms proposed to
explain the observed damping, e.g., footpoint leakage
\citep{bergbruyne95,depontmart01}, phase mixing \citep{heypri83,roberts02,ofasch02}
and resonant absorption \citep{rudrob02,goossetal02} and more recently loop cooling
\citep{mortonerd09}. Thus far, resonant absorption, caused by plasma inhomogeneity
in the direction transverse to the magnetic field, has proved the most likely
candidate in explaining the observed short damping times in coronal loops
\citep[see][for review]{goossens08}. Consistent seismological studies based on
resonant absorption using the observed values of periods and damping times of
standing kink waves were carried out by \citet{arreguietal07} and
\citet{goossetal08}. These two studies show that, at least for a collection of 11
loops, resonant absorption provides an explanation of the observed periods and
damping times. The observational signatures of the alternative cooling loop damping
mechanism proposed by \citet{mortonerd09}, differ from resonant absorption by the
fact that frequency changes as a function of time (the resonant damping theory
developed so far is restricted to static equilibria and therefore frequency is
expected to be constant in time). Extensive MHD modelling has also shown that
resonant absorption is the most likely explanation for the damping of transverse
oscillations in prominence fine structures
\citep[see][]{arretal08,soleretal09b,soleretal09c,arrball10}.



If indeed, coronal loop kink oscillations are being attenuated by the process of
resonant absorption, one can exploit this to estimate the transverse plasma
inhomogeneity length scales using observed frequencies and damping times
\citep{goossensetal06}. The original equilibrium models by e.g., \citet{rudrob02},
to study resonant absorption consisted of monolithic loop structures. However,
there has been some observational evidence by e.g., \citet{aschw05} that coronal
loops are composed of many different strands, possibly at different temperatures
and densities. To model this loop multi-thread structure, there was an initial
numerical study by \citet{terrarretal08} and it was found that the process of
resonant absorption was still an efficient damping mechanism in more complex and
realistically structured loop models. Further study into the properties of standing
kink waves was also undertaken relating plasma inhomogeneity in the direction of
the magnetic field caused by e.g., density
\citep{diazetal04,andriesetal05,arretal05,dymova06,erdverth07,verthetal07} or
magnetic \citep{vertherd08,ruderetal08} stratification. It was shown that
eigenfrequencies and eigenfunctions of coronal loops with longitudinal
stratification, were altered in such a way that one could determine, e.g., the
coronal density scale height by estimating the ratio of the fundamental mode to
that of higher overtones, a technique first proposed by \citet{andriesetal05a} and
later developed by \citet{verthetal08} to correct for magnetic stratification.

All this theoretical development to describe kink waves in coronal loops with
realistic plasma stratification in the transverse and longitudinal direction was
restricted to the study of standing waves, since this was what was detected in
TRACE data. However, recently it has come to light that there are also ubiquitous
small amplitude propagating transversal MHD waves in the solar atmosphere. These
were first observed by the novel Coronal Multi-Channel Polarimeter (CoMP)
instrument \citep{tomczetal07}. Moreover, \citet{tomcmac09} have been able to
separate outward and inward propagating wave power. It was found that the outward
power was greater than the inward power by about a factor of two and this can only
be explained if the waves are damped in situ \citep[see also][]{pascoeetal10}. The
reason we could not detect these propagating waves previously with TRACE is that
the amplitudes are of the order 50 km, while the TRACE resolution is only about 800
km. \citet{tomczetal07} originally interpreted these wave as Alfv\'{e}n waves,
however, \citet{vandetal08b} subsequently argued that the observed waves were
actually more consistent with the propagating kink mode. Although both modes are
dominated by the restoring force of magnetic tension, in the geometry of a solar
flux tube, e.g., a magnetic cylinder, a pure Alfv\'{e}n wave is strictly torsional
with no transverse component and therefore completely incompressible. On the other
hand, a kink wave propagating in a flux tube has a transverse perturbation
component and is weakly compressible, at least in the linear regime \citep[see
e.g.,][]{goossensetal09}.

In the present paper we shall restrict the study to investigating the effect of
transverse plasma inhomogeneity on the propagating kink mode. Although the two
problems of standing and propagating kink waves are closely related, since a
standing wave is a superposition of two propagating waves, there are some
differences that need to be considered. The standing transverse oscillation is the
result of an initial value problem, i.e., an initial disturbance in the solar
corona, e.g., a CME or a flare, induces the oscillations of the loops at their
natural frequencies or eigenmodes. Alternately, the transverse travelling waves
have a forced nature since the photosphere acts as a driver. The frequencies of the
kink waves observed by CoMP show a peak around 5 min, indicating a $p$-mode driven
photospheric origin. The spatial scale of the driver at the base of the tube is
also important in order to excite kink oscillations. From the properties of MHD
waves in a flux tube we know that a purely incompressible excitation excites purely
incompressible Alfv\'en waves if the driver is strictly localised inside or outside
the tube (assuming a homogeneous loop model). On the other hand, an excitation
located both inside and outside the tube invariably excites transverse oscillations
since an almost incompressible surface wave between the two media will be
established, i.e., the kink mode.

\section{Waveguide model}\label{model}

To understand the effect of radial inhomogeneity on propagating kink waves we
consider the relatively simple equilibrium model of a cylindrical axi-symmetric
flux tube of radius $R$ with a constant axial magnetic field $B_z$ and with a
density contrast of $\rho_{\mathrm i}/\rho_{\mathrm e}$. The subindex ``${\mathrm
i}$'' and ``${\mathrm e}$'' refer to the internal and external part of the tube,
respectively. It is also assumed that the tube has a smooth variation of density
across the waveguide boundary (located at $r=R$) with a characteristic spatial
scale $l$. For simplicity a sinusoidal density profile connecting the internal and
external part of the tube is implemented \citep[see
e.g.,][]{rudrob02,goossetal02,terretal06b}. The role of the inhomogeneity at the
loop boundary is crucial since this is where the process of resonant absorption
invariably takes place. This basically means that the transverse displacement of
the whole tube is converted into azimuthal motions localised at the tube boundary.
Due to this energy conversion, the transverse motion is attenuated, while at the
same time small scales are created in the nonuniform layer due to phase mixing.
Phase mixing causes a cascade of energy to smaller length scales, where the
dissipation becomes more efficient. The reader is referred to \citet{goossens08,
ruderd09,terradas09} and references therein for further details about this robust
damping mechanism. Recent studies by \citet{soleretal09b} in the context of
modelling kink oscillations observed in solar prominences (complex coronal magnetic
structures with relatively cool and dense plasma), show that the process of
resonant absorption still survives even when the plasma is partially ionised.

\section{Spatial damping in the TT approximation}\label{TT}

Before we embark on the analysis of MHD waves  in the thin tube (TT)
approximation, let us explain what this approximation actually is. In
what follows we shall consider both standing and propagating waves. A standing
wave is the superposition of two propagating waves travelling with the same frequency and wavenumber but in opposite directions. For the
standing wave problem the axial wavelength (or the axial wavenumber $k_z$) is
specified and the corresponding frequency is determined.  A TT in this
case means that the axial wavelength is much longer than the radius of the tube
so that $k_zR \ll 1$. Hence the TT approximation for standing waves is
the long wavelength approximation. For propagating waves the frequency $\omega$
is specified and the corresponding wavelength is determined. In this situation,
a TT means that during one period, defined by the frequency $\omega$, a
signal travelling at the Alfv\'{e}n speed can cross the waveguide in the
radial direction many times or $\omega/(v_{\mathrm A}/R) \ll 1$. Hence for
propagating waves, the TT approximation is the low frequency
approximation. The benefit of the TT approximation is that it enables us to
obtain simple mathematical expressions which are very accurate, allowing us to
gain a physical insight into the problem.

\subsection{Homogeneous magnetic cylinder}  For a homogeneous magnetic cylinder
(i.e., no inhomogeneous layer, $l=0$) the dispersion relation for the kink mode
($m=1$) and the fluting modes ($m>1$) in the TT approximation is
\citep[see][]{goossensetal09}
\begin{equation}\label{TTdisper}
 \rho_{\mathrm i} (\omega^2 - \omega_{\mathrm
{Ai}}^2)+\rho_{\mathrm e} (\omega^2 - \omega_{\mathrm {Ae}}^2)=0 , \label{F}
\end{equation}
where
\begin{eqnarray}
\omega_{\mathrm A}^2 & = k_z^2 v_{\mathrm A}^2,
\;\; v_{\mathrm A}^2 = B_z^2/ \left(\mu \, \rho\right).
\label{OmegaAC}
\end{eqnarray}

\noindent The dispersion relation Eq.~(\ref{TTdisper}) specifies a functional
dependence on frequency $\omega$ and wavenumber $k_z$. This can be studied for
the cases of either standing or propagating waves. For the propagating wave
study, the
wavenumber $k_z$ is specified and the dispersion relation is solved for
frequency $\omega$, leading to the well known result
\begin{equation}
\omega^2=\frac{\rho_{\mathrm i} \omega_{\mathrm {Ai}}^2+\rho_{\mathrm e}\omega_{\mathrm
{Ae}}^2}{\rho_{\mathrm i}+\rho_{\mathrm e}}\equiv \omega_*^2,
\end{equation}
where $\omega_*$ is real. For equal magnetic field strength inside and outside
the cylinder this expression can be further simplified to
\begin{equation}\label{kinksimp}
\omega^2=\frac{2B_z^2}{\mu\,(\rho_{\mathrm i}+\rho_{\mathrm e})}k_z^2=
\frac{2\rho_{\mathrm i}}{\rho_{\mathrm i}+\rho_{\mathrm e}}v_{\mathrm {Ai}}^2 k_z^2
\equiv\omega_*^2.
\end{equation}
For propagating waves we consider waves generated at a given location with a real
frequency $\omega_*$ and solve the dispersion relation for $k_z$, resulting in
\begin{equation}\label{TTkz}
k_z^2  =  \frac{\mu\,(\rho_{\mathrm i} +\rho_{\mathrm e})}{2 B_z^2}
\, \omega_*^2=\frac{\rho_{\mathrm i} +\rho_{\mathrm e}}{2 \rho_{\mathrm i}}
\, \frac{\omega_*^2} {v_{\mathrm {Ai}}^2} \equiv k_*^2,
\label{Freqkink}
\end{equation}
where $k_*$ is also real. The solution given by
Eq.~(\ref{TTkz}) corresponds to the well known undamped kink wave (and the
fluting modes). To derive Eq.~(\ref{TTdisper}) it is implicitly assumed that
$\omega R/v_{\mathrm A}\ll 1$ but  since not all
frequencies will satisfy these conditions later we present results for
any frequency of driver without this restriction.

\subsection{Inhomogeneous magnetic cylinder}\label{TTnonunif}

The thin boundary (TB) approximation means that the non-uniformity is confined to
$[R-l/2,R+l/2]$, with $l/R \ll 1$, so that the non-uniform layer coincides with
the dissipative layer. This approximation results in the mathematical
simplification that MHD waves can be described by solutions for uniform plasmas
that are connected over the dissipative layer by jump conditions \citep[see
e.g.,][]{sakuraietal91,goossensetal95,tirrygoss96}. The use of the TB
approximation has been applied in e.g.,
\citet{goossensetal06}, \citet{goossens08} and \citet{goossensetal09}. The inclusion of the effect of
an inhomogeneous layer is reasonably simple in the case of the thin boundary
approximation and results in the following dispersion relation,
\begin{eqnarray}
\rho_{\mathrm i} (\omega^2 - \omega_{\mathrm {Ai}}^2)&+&\rho_{\mathrm e} (\omega^2
- \omega_{\mathrm {Ae}}^2)=\nonumber \\
& & i \pi \frac{ m/ r_{\mathrm A}}{\rho(r_{\mathrm A}) \left|\Delta_{\mathrm
A}\right|} \rho_{\mathrm i} \left(\omega^2 - \omega_{\mathrm {Ai}}^2\right)
\rho_{\mathrm e} \left(\omega^2 - \omega_{\mathrm {Ae}}^2\right). \label{G}
\end{eqnarray}
Here $r_{\mathrm A}$ denotes the position of the Alfv\'{e}n resonance. In
the TB approximation it is natural to adopt $r_{\mathrm A}=R$ since $l/R \ll 1$
and $r_{\mathrm A}\in]R-l/2,R+l/2[$. Use of the jump condition is not restricted to the
thin non-uniform layer as can be seen from e.g.,
\citet{tirrygoss96,tirryetal97,tirryetal98}. However, this condition requires
numerical integration of the ideal MHD equations in a non-uniform plasma up to
the dissipative layer. In the present paper we do not intend to use numerical
integration of the ideal MHD equations relating to thick boundaries. However, we
shall use the results obtained with the TB approximation for thick boundaries for a
comparison with the results of a full numerical calculation in Section~\ref{RES}.


Note that the effect of resonance is contained in the right hand side of
Eq.~(\ref{G}). Again we can view the dispersion relation Eq.~(\ref{G}) as a
relation for either standing or propagating waves. In the case of standing waves
the wavenumber is real ($k_z=k_*$) and the frequency is complex. This case has
been considered previously by e.g., \citet{goossensetal92,goossensetal09}.
Let us now focus on propagating waves with a given real frequency
($\omega=\omega_*$) and complex wavenumber. The imaginary part of the wavenumber
indicates that the wave, as it propagates, is damped due to resonant absorption.
Now we assume that
\begin{equation}\label{kcomplex} k_z =  k_{\mathrm R}+i\,
k_{\mathrm I}. \end{equation}
The purely imaginary term in Eq.~(\ref{kcomplex})
reflects the damping imposed on the wave and since the damping is in the spatial
domain the wavenumber is now complex. If we approximate $k_z^2$ by $k_{\mathrm
R}^2+2 i k_{\mathrm R} k_{\mathrm I}$ (we assume weak damping, i.e., $k_{\mathrm I}\ll
k_{\mathrm R}$) we have that $ k_{\mathrm R} \approx k_*$ (given by
Eq.~[\ref{Freqkink}]) and after some algebra we find
\begin{equation}
\frac{k_{\mathrm I}}{k_*}=\frac{\pi}{8} \frac{ m}{R}\frac{1}{\rho(r_{\mathrm A})
\left|
\Delta_{\mathrm A}\right|} \frac{\left(\rho_{\mathrm i} - \rho_{\mathrm
e}\right)^2}{\left(\rho_{\mathrm i} + \rho_{\mathrm e}\right)}\,\omega_*^2.
\end{equation}
By the fact that,
 \begin{equation} \rho(r_{\mathrm A}) \left|
\Delta_{\mathrm A}\right| =\omega^2\left|\frac{ d \rho}{ dr}
\right|_{r_{\mathrm A}}, \end{equation}
we finally obtain the following
expression
\begin{equation}\label{kioks} \frac{k_{\mathrm I}}{k_*}=\frac{\pi}{8}
\frac{ m}{R} \frac{\left(\rho_{\mathrm i} - \rho_{\mathrm
e}\right)^2}{\rho_{\mathrm i} + \rho_{\mathrm e}}\frac{ 1}{ \left|d \rho/dr
\right|_{r_{\mathrm A}}}. \end{equation}
 \noindent Because $k_z$ is complex
we define the damping length as $L_{\mathrm D}=1/k_{\mathrm I}$ while the
wavelength is simply $\lambda=2\pi/k_*$. A useful quantity is the the damping per wavelength
which is
\begin{equation}\label{ldolambtt} \frac{ L_{\mathrm D}}{ \lambda}
=\frac{4}{\pi^2} \frac{ R}{m} \frac{\rho_{\mathrm i} + \rho_{\mathrm
e}}{\left(\rho_{\mathrm i} - \rho_{\mathrm e}\right)^2}{ \left|\frac{ d
\rho}{ dr} \right|_{r_{\mathrm A}}}. \end{equation}
For a sinusoidal density
profile it can be shown that
\begin{equation}\label{sinus} { \left|\frac{ d \rho}{
dr} \right|_{r_{\mathrm A}}}=\frac{\pi}{2} \frac{\rho_{\mathrm i} -
\rho_{\mathrm e}}{l}, \end{equation} and Eq.~(\ref{ldolambtt}) reduces to
\begin{equation}\label{damplength} \frac{ L_{\mathrm D}}{ \lambda} =\frac{2}{\pi}
\frac{1}{m} \frac{ R}{l} \frac{\rho_{\mathrm i} + \rho_{\mathrm e}}{\rho_{\mathrm
i} - \rho_{\mathrm e}}. \end{equation}
Equation~(\ref{damplength}) clearly shows the
dependence of the damping per wavelength on the thickness of the layer. The
wider the layer, the stronger the spatial attenuation of the wave. This is not
surprising, since we can relate this result to the expression for the temporal
damping, i.e., wavenumber assumed real ($k_z=k_*$) and frequency complex. The
well known formula in the case of the temporal damping for a standing wave is
\begin{equation}\label{damptime} \frac{ \tau_{\mathrm D}}{  P} = \frac{2}{\pi}
\frac{1}{m}\frac{ R}{l} \frac{\rho_{\mathrm i} + \rho_{\mathrm e}}{\rho_{\mathrm
i} - \rho_{\mathrm e}}. \end{equation}

\noindent If we compare this expression with Eq.~(\ref{damplength}) we note that
the damping per wavelength for propagating waves and the damping per period for
standing waves  are exactly the same. Hence, in the TT approximation spatial and
temporal damping are completely equivalent. From Eqs.~(\ref{damplength}) and
(\ref{damptime}) we obtain the simple result
\begin{equation}
\frac{L_{\mathrm D}}{\lambda}=\frac{\tau_{\mathrm D}}{P}.
\label{equiv}
\end{equation}
Propagating kink waves were recently studied by \citet{farahanietal09} in the long
wavelength limit in X-ray jets in the solar atmosphere. These authors computed the
ratio of the damping time to the period for standing waves and used this quantity
to discuss the spatial damping of propagating waves. The TT approximation result
presented in Eq.~(\ref{equiv}), therefore validates the use of the damping
expression derived for kink standing waves by \citet{farahanietal09} to interpret the
attenuation of propagating kink waves. The TT damping relations given by
Eqs.~(\ref{damplength}) and (\ref{damptime}) have further significant consequences.
For standing waves we rewrite Eq.~(\ref{damptime}) as
\begin{equation}
\frac{\tau_{\mathrm D}}{P}=\frac{\xi_E}{m},
\label{damptimefE}
\end{equation}
where
\begin{equation}
\xi_E=\frac{2}{\pi}\frac{R}{l}\frac{\rho_{\mathrm i}+\rho_{\mathrm
e}}{\rho_{\mathrm i}-\rho_{\mathrm e}},
\label{xidef}
\end{equation}
which only depends on the parameters of the equilibrium model, not on the
particular type of MHD wave mode defined by the value of $m$ (note that for the
kink mode $m=1$). The period is defined by
\begin{equation}
P=\frac{2 \pi}{\omega},
\end{equation}
and for standing waves $\omega$ is related to the wavenumber $k_z$ by the
dispersion relation given by Eq.~(\ref{kinksimp}), i.e.,
\begin{equation}
\omega=n \, v_{\mathrm {Ai}} \frac{\pi}{L} \sqrt{\frac{2 \rho_{\mathrm i}}{\rho_{\mathrm
i}+\rho_{\mathrm e}}},
\end{equation}
where $v_{\mathrm {Ai}}$ is the internal Alfv\'{e}n speed, $n=1,\,2,\,
3,\,\ldots$ is the longitudinal mode number and $L$ is the total length of the
waveguide, e.g., a coronal loop (we have used that $k_z=n \pi/L$).
Equation~(\ref{damptimefE}) can then be written as
\begin{equation}
\frac{\tau_{\mathrm D}}{\tau_{\mathrm {Ai}}}=\frac{2\xi_E}{m}\sqrt{\frac{\rho_{\mathrm
i}+\rho_{\mathrm e}}{2\rho_{\mathrm i}}}\, \frac{1}{n},
\label{DTn}
\end{equation}
where $\tau_{\mathrm {Ai}}=L/v_{\mathrm {Ai}}$ is the Alfv\'{e}n transit time in
the longitudinal direction. Equation~(\ref{DTn}) has interesting consequences.
Firstly, it indicates that fluting modes ($m>1$) have shorter damping times than
the kink mode ($m=1$). Secondly, it shows that the damping time for a standing
wave is inversely proportional to the longitudinal mode number $n$, i.e., the
damping time is inversely proportional to the wavelength of the standing wave, so
that higher overtones (with shorter periods) are damped faster than low order
overtones, e.g, the fundamental mode. Fortunately, there have been some signatures of
overtones in coronal loop standing kink waves detected in TRACE data
\citep[see e.g.][]{verwetal04,demoortbrady07,verthetal08,vandetal09}. For a
particularly clear example of the first overtone damping before the fundamental
mode, that could be explained by resonant absorption attenuating the higher
harmonic faster, see the Morlet wavelet transform in Figure~5 of
\citet{verwetal04}.

Now considering the spatial damping of propagating waves we can also
write Eq.~(\ref{damplength}) as
\begin{equation}
\frac{L_{\mathrm D}}{\lambda}=\frac{\xi_E}{m}.
\label{DTm}
\end{equation}
The wavelength is defined as $\lambda=2 \pi/k_z$ and is related to the frequency
by the dispersion relation in Eq.~(\ref{Freqkink}). Equation~(\ref{DTm}) can then be rewritten in terms of the flux tube radius as
\begin{equation}
\frac{L_{\mathrm D}}{R}=2\pi\frac{v_{\mathrm {Ai}}}{\omega R}
\frac{\xi_E}{m}\sqrt{\frac{2 \rho_{\mathrm i}}{\rho_{\mathrm
i}+\rho_{\mathrm e}}},
\label{dlE}
\end{equation}
where $\omega R/v_{\mathrm {Ai}}$ is a dimensionless frequency (note that the TT
approximation means
$\omega R/v_{\mathrm {Ai}}\ll 1$). Denoting this frequency as
\begin{equation}
f=\frac{\omega R}{v_{\mathrm {Ai}}},
\label{domega}
\end{equation}
then
\begin{equation}\label{ldoRtt}
\frac{L_{\mathrm D}}{R}=2\pi\frac{\xi_E}{m}\sqrt{\frac{2 \rho_{\mathrm i}}{\rho_{\mathrm
i}+\rho_{\mathrm e}}}\, \frac{1}{f}.
\end{equation}
We shall refer to the relation between the damping length and frequency defined in Eq.~(\ref{ldoRtt}) as the TGV relation for propagating kink waves. In what follows, it will become clear that the TGV relation is actually a very good approximation of the dependency
of damping length on frequency for all relevant frequencies as illustrated in Fig. 2. The expression given in Eq.~(\ref{ldoRtt}) has important consequences, as it shows that $L_{\mathrm D}/R$ is
inversely proportional to $f$ for propagating waves, i.e., the damping length is
inversely proportional to frequency, so that high frequency waves are damped in
shorter spatial scales than their lower frequency counterparts. This means that for driven waves propagating upwards from the photosphere
each frequency has a different penetration height into the solar corona.
Therefore, resonant absorption provides a natural filtering mechanism for a
broadband disturbances, e.g., like those observed by \citet{tomcmac09}, with
lower frequency waves being least affected by the damping process, propagating to higher heights in the solar corona.

\begin{table*}
\caption{Estimated range of $k_z R$ in observed propagating (P) and standing (S) MHD wave modes.}
\centering 
\begin{tabular}{l l l l l l} 
\hline\hline 
Reference& MHD wave mode&Width of & Wavelength& $k_z R$\\
& interpretation of authors& waveguide&(Mm) & \\
 & &(Mm) & &\\
 \hline 
 &&&&\\
1&Kink (S)&$5.9-11.5$&$328-552$&$0.03-0.11$\\
& & & &\\
2&Alfv\'{e}n (P)&0.2 & $> 4.0$ &$<0.16$\\
& & & &\\
3&Kink or sausage waves (S)& 0.4 &$2.5-3.1$&$0.41-0.5$\\
&&&&\\
4&Torsional Alfv\'{e}n (P)&2.0 &$4.5-6.0$&$1.04-1.39$\\
& & & & &\\
5&Kink (P)&$0.2-1.0$& $3.4-12.9$ & $0.03-0.9$\\
& & & &\\
6&Kink (P)&$0.43-0.66$& $> 250$ & $< 0.12$\\
& & & &\\
7&Alfv\'{e}n (P)&$9.0$&$> 393$&$<0.07$\\
&& & &\\
8&Kink or Alfv\'{e}n waves (P)&$9.0$&$>180$&$<0.16$\\
&&&&\\
\hline 
\end{tabular}
\tablebib{(1) \citet{aschetal02}; (2) \citet{depontetal07}; (3)
\citet{fujimuratsuneta09}; (4) \citet{jessetal09}; (5) \citet{linetal09}; (6)
\citet{okamoto07}; (7) \citet{tomczetal07}; (8) \citet{tomcmac09}.}
\label{tab1} 
\end{table*}

\section{Spatial damping beyond the TT approximation}\label{noTT}

The TT approximation is useful because it makes the dispersion relation
analytically solvable. In Table~\ref{tab1} we list observed estimates of $k_z R$
for various MHD wave modes and it can be seen that for many solar atmospheric
waves, e.g., standing kink waves in coronal loops observed by \citet{aschetal02},
the TT approximation is reasonably valid. However, for kink waves in filament
threads \citep{linetal09} or torsional Alfv\'{e}n waves in chromospheric magnetic
bright points \citep{jessetal09} may have values of $k_z R\approx 1$. To address
the kink waves observed in this regime, it is relatively straight forward to relax
the TT approximation and to take into account the effect of the finite tube radius
on damping.

\subsection{Homogeneous magnetic cylinder}

For a homogeneous magnetic cylinder the dispersion relation, taking into account a
finite radius, is
\begin{equation} \rho_{\mathrm i} (\omega^2 - \omega_{\mathrm
{Ai}}^2)+\rho_{\mathrm e} (\omega^2 - \omega_{\mathrm {Ae}}^2)F(\omega,k_z)=0 ,
\label{noTTdisper}
\end{equation}
where
\begin{eqnarray}
F(\omega,k_z)= -\frac{k_{\mathrm i}}{k_{\mathrm e}}
\frac{J'_m(k_\mathrm i R)K_m(k_\mathrm e R)}{J_m(k_\mathrm i R)K'_m(k_\mathrm
e R)},
\label{ffunct}
\end{eqnarray}
\begin{eqnarray}
k^2_{\mathrm i}= \frac{\omega^2-\omega_{\mathrm {Ai}}^2}{v^2_{\mathrm {Ai}}},\;
k^2_{\mathrm e}= -\frac{\omega^2-\omega_{\mathrm {Ae}}^2}{v^2_{\mathrm {Ae}}}.
 \label{ke}
\end{eqnarray}

\noindent Note that the function $F$ depends both on frequency $\omega$ and wave
number $k_z$, taking into account that the tube has a finite width. In the limit
of a TT, $F\equiv1$ (the Bessel functions are approximated by their small
arguments expressions) and we recover the dispersion relation given by
Eq.~(\ref{TTdisper}). For a standing wave $k_z$ is fixed and
Eq.~(\ref{noTTdisper}) is solved for $\omega$. For a propagating wave $\omega$ is
fixed and Eq.~(\ref{noTTdisper}) is solved for $k_z$. The solution to
Eq.~(\ref{noTTdisper}) corresponds again to the undamped kink wave. To solve this
equation analytically for a finite width tube one would have to use additional
terms in the asymptotic expansions of the Bessel functions, complicating matters.
To avoid these cumbersome calculations we simply solve Eq.~(\ref{noTTdisper})
numerically, with the solution differing from the solution given by
Eq.~(\ref{TTkz}) outwith the TT regime.

\subsection{Inhomogeneous magnetic cylinder}
When we add a thin non-uniform layer we obtain the complex dispersion relation
\begin{eqnarray}\rho_{\mathrm i} (\omega^2&-&\omega_{\mathrm
{Ai}}^2)+\rho_{\mathrm e} (\omega^2-\omega_{\mathrm {Ae}}^2)\,F(\omega,k_z)=\nonumber\\
&&i \pi \frac{ m/ R}{\rho(r_{\mathrm A}) |\Delta_{\mathrm A}| }
\rho_{\mathrm i} \left(\omega^2-
\omega_{\mathrm {Ai}}^2\right) \rho_{\mathrm e} \left(\omega^2 -
\omega_{\mathrm {Ae}}^2\right) G(\omega, k_z), \label{noTTdisperres}
\end{eqnarray}
where
\begin{eqnarray}
G(\omega, k_z)= -\frac{m}{R}\frac{1}{k_\mathrm e}
\frac{K_m(k_\mathrm e R)}{K'_m(k_\mathrm e R)}.
\label{gfunct}
\end{eqnarray}
\noindent For the TT we can again use the asymptotic expansions of the Bessel
function so that $G\equiv1$, hence in the TT limit, Eq.~(\ref{noTTdisperres}) is
simplified to Eq.~(\ref{G}). Equation~(\ref{noTTdisperres}) can be solved for a
complex frequency $\omega$ with a specified real wavenumber $k_*$ or for a
complex wavenumber $k_z$ for a specified real frequency $\omega_*$. By solving
Eq.~(\ref{noTTdisperres}) in the case of a propagating wave with real frequency
$\omega_*$ we find that the real part of complex $k_z$ is $k_*$ (we get again
Eq.~[\ref{noTTdisper}], note that now $k_*$ is different from Eq.~[\ref{G}]),
while the imaginary part is given by
\begin{equation}\label{kioksnoTT}
\frac{k_{\mathrm I}}{k_*}=\frac{T}{N} ,
\end{equation}
where
\begin{eqnarray}\label{a}
T=\frac{\pi}{2} \frac{
m}{R} \frac{\left(\rho_{\mathrm
i} - \rho_{\mathrm e}\right)^2}{\left[\rho_{\mathrm i} + \rho_{\mathrm
e}F(\omega_*, k_*)\right ]\left[1 + F(\omega_*, k_*)\right]^2}\times\nonumber \\
\frac{F(\omega_*, k_*)\,G(\omega_*,k_*)}{ |
d \rho/dr |_{r_{\mathrm A}}},
\end{eqnarray}
and
\begin{equation}\label{b}
N=1+ \frac{k_*}{2} \frac{\rho_{\mathrm i} -
\rho_{\mathrm e}}{\left[\rho_{\mathrm i} + \rho_{\mathrm e}F(\omega_*, k_*)\right
]\left[1 + F(\omega_*,k_*)\right]}\left.\frac{\partial F}{\partial
k}\right|_{(\omega_*, k_*)}.
\end{equation}
\noindent The expressions are slightly more complicated than in the
TT approximation (see Eq.~[\ref{kioks}]), but once $k_*$ is known the different terms can
be easily calculated. The corresponding damping per
wavelength is
\begin{eqnarray}\label{ldolambnoTT}
\frac{L_{\mathrm D}}{ \lambda} =\frac{1}{2\pi}\frac{N}{T}.
\end{eqnarray}
\noindent In the TT limit, i.e., when $F(\omega_*,k_*)$ and $G(\omega_*, k_*)$ $\rightarrow 1$ and $\partial F/\partial k \rightarrow 0$, we recover Eq.~(\ref{ldolambtt}). For a sinusoidal density
profile we simply have to make use of Eq.~(\ref{sinus}) in Eq.~(\ref{a}).
 \noindent Now let us define
\begin{equation}
\widetilde{N}=1-\frac{\omega_*}{2}\frac{\rho_{\mathrm
i} - \rho_{\mathrm e}}{\left[\rho_{\mathrm i} + \rho_{\mathrm
e}F(\omega_*, k_*)\right ]\left[1 + F(\omega_*, k_*)\right]} \left.\frac{\partial
F}{\partial \omega}\right|_{(\omega_*, k_*)}.
\end{equation}
As in the previous Section we compare the damping per wavelength for propagating
waves with the damping per period, which in the non-TT approximation is given by
\begin{eqnarray}
\frac{\tau_{\mathrm D}}{ P}=\frac{1}{2 \pi}\frac{\widetilde{N}}{T}.
\label{tDP}
\end{eqnarray}
\noindent By Eqs. (\ref{ldolambnoTT}) and (\ref{tDP}), there is a clear correspondence between the temporal and the spatial
damping beyond the TT approximation. Furthermore, a general expression that relates the temporal damping of
standing waves and the spatial damping of propagating waves can be derived
following the analysis of Appendix A in \citet{tagger95}. We can write the
complex dispersion relation given by Eq.~(\ref{noTTdisperres}) as
\begin{eqnarray}\label{ldtd1}
D_{\mathrm R}(\omega,k)+i D_{\mathrm I}(\omega,k)=0.
\end{eqnarray}
To make it clear, in Eq.~(\ref{ldtd1}), $D_{\mathrm R}$ is equal to the LHS of
Eq.~(\ref{noTTdisperres}) and $D_{\mathrm I}$ equal to minus the RHS of the
equation. In the case of spatial damping, if $\omega_*$ and $k_*$ are the
solutions of $D_{\mathrm R}(\omega,k)=0$ then it is easy to see by making a Taylor
expansion around the solution that
\begin{eqnarray}\label{ldtd2}
k_{\mathrm I}=-\left.\frac{D_{\mathrm I}}{\partial D_{\mathrm
R}/\partial k}\right|_{(\omega_*,k_*)},
\end{eqnarray}
while for temporal damping
\begin{eqnarray}\label{ldtd3}
\omega_{\mathrm I}=-\left.\frac{D_{\mathrm I}}{\partial D_{\mathrm R}/\partial \omega}\right|_{(\omega_*,k_*)}.
\end{eqnarray}
Combining these two expressions we find that
\begin{eqnarray}\label{ldtd4}
\frac{L_{\mathrm D}}{\lambda}=\frac{k_*}{\omega_*}\frac{{\partial D_{\mathrm R}/\partial
k}}{{\partial D_{\mathrm R}/\partial \omega}}\frac{\tau_{\mathrm D}}{ P},
\end{eqnarray}
which is simply \citep[see also][]{pascoeetal10}
\begin{eqnarray}\label{ldtdgr}
\frac{L_{\mathrm D}}{\lambda}=\frac{v_{\mathrm {gr}}}{v_{\mathrm {ph}}}\frac{\tau_{\mathrm D}}{ P}.
\end{eqnarray}
\noindent This result is valid when the damping is not too strong \citep[see
e.g.,][]{tagger95}, an assumption that we have already made in the derivation of
the damping per wavelength ($k_{\mathrm I}\ll k_{\mathrm R}$). In the TT approximation kink waves are weakly
dispersive so that $v_{\mathrm {gr}}\approx v_{\mathrm {ph}}$ hence, in agreement with
the results found in Section~\ref{TT}, the damping per wavelength is exactly
the same as the damping per period but in the regime when the TT approximation is
not applicable the group speed can differ from the phase speed.

From the previous results we can also derive alternative formulae for the damping
per wavelength and damping per period in terms of the imaginary part of the dispersion relation,
\begin{eqnarray}\label{ldtdalt} & &\frac{L_{\mathrm
D}}{\lambda}=\frac{k_*}{2\pi}\frac{1}{D_{\mathrm I}(\omega_*, k_*)}\times \nonumber \\
& &\left\{-2 k_*\left[\rho_{\mathrm i}v^2_{\mathrm {Ai}}+ \rho_{\mathrm
e}v^2_{\mathrm {Ae}} F(\omega_*,k_*)\right] + \rho_{\mathrm
e}(\omega_*^2-k^2_*v^2_{\mathrm {Ae}})\left.\frac{\partial F}{\partial
k}\right|_{(\omega_*,k_*)}\right\}, \nonumber \\
\end{eqnarray}
\begin{eqnarray}\label{ldtdm} \frac{\tau_{\mathrm
D}}{P}&=&\frac{\omega_*}{2\pi}\frac{1}{D_{\mathrm I}(\omega_*, k_*)}\times
\nonumber \\ & &\left\{2
\omega_*\left[\rho_{\mathrm i}+\rho_{\mathrm e}
F(\omega_*,k_*)\right]+\rho_{\mathrm e}(\omega_*^2-k^2_*v^2_{\mathrm
{Ae}})\left.\frac{\partial F}{\partial \omega}\right|_{(\omega_*,k_*)}\right\}. \nonumber \\ \end{eqnarray}
Using Eq.~(\ref{ldtdgr}) we can identify in these expressions the terms
related to the phase and group speed.

\begin{figure}[!ht]
\center{\includegraphics[width=9.25cm]{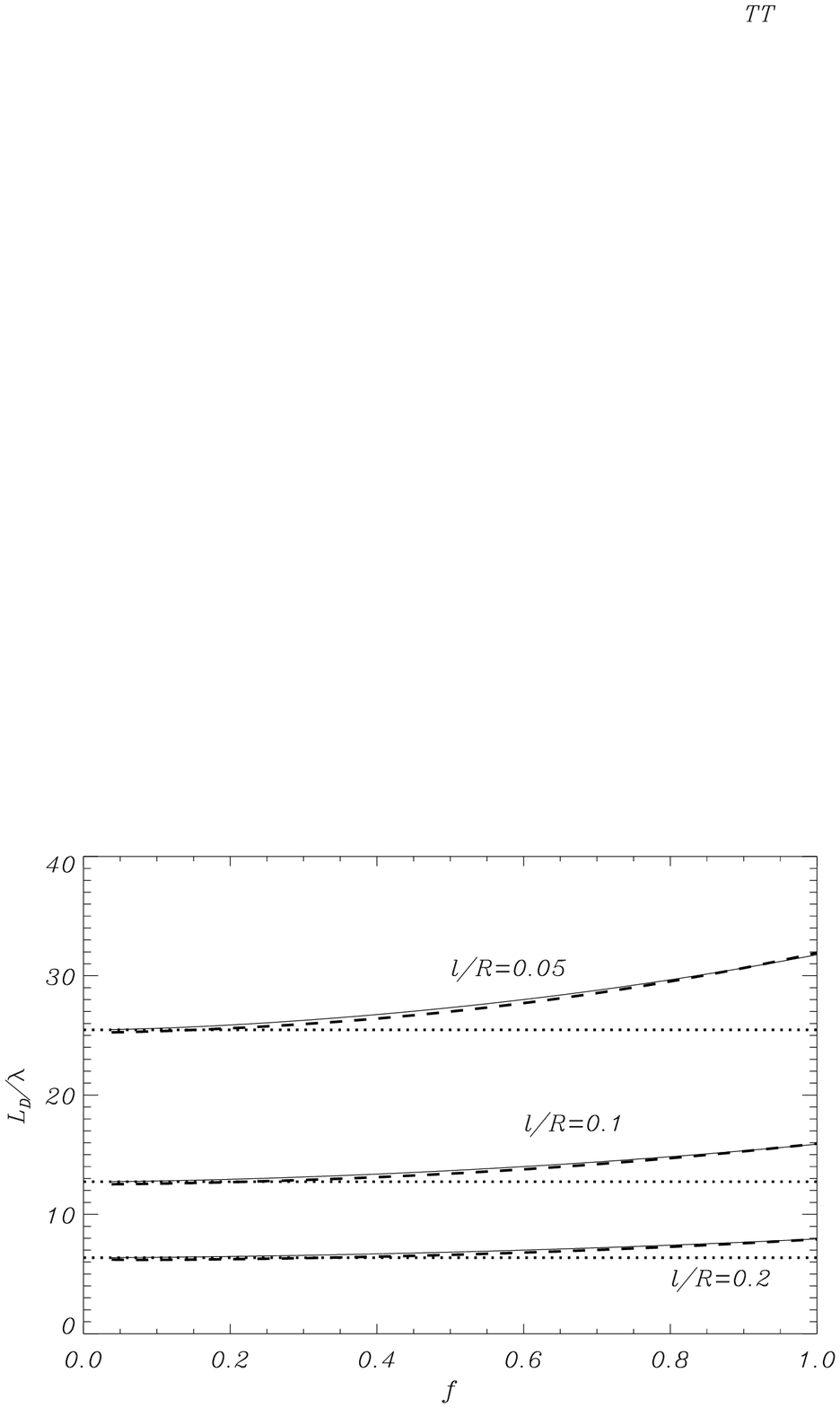}} \caption{\small
Damping per wavelength as a function of the dimensionless
frequency ($f=\omega R/v_{\mathrm {Ai}}$) for three different widths
of the inhomogeneous layers. The solid line corresponds to the analytical results, the dashed
line represents the full numerical solution of the resistive eigenvalue problem.
The dotted line corresponds to the TT approximation, valid when $f\rightarrow 0$. In this plot
$\rho_{\mathrm i}/\rho_{\mathrm e}=3$. } \label{Ldolambda} \end{figure}

\begin{figure}[!ht]
\center{\includegraphics[width=9.25cm]{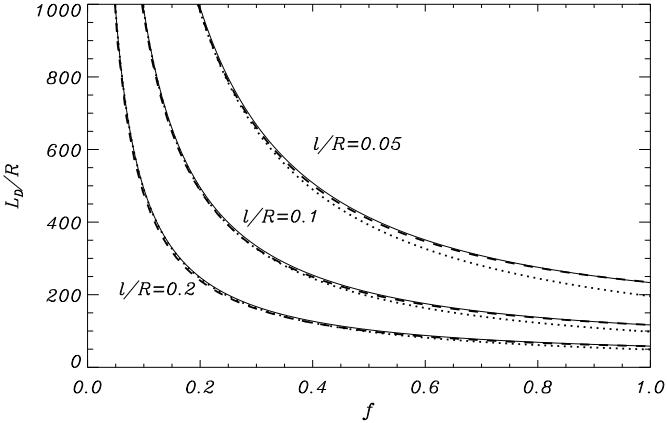}} \caption{\small
Damping length normalised to the loop radius as a function of the dimensionless
frequency ($f=\omega R/v_{\mathrm {Ai}}$) for three different widths of the
inhomogeneous layers. The solid line corresponds to the analytical results, the dashed line
represents the full numerical solution of the resistive eigenvalue problem. The
dotted line corresponds to the TT approximation calculated using
Eq.~(\ref{ldoRtt}), valid when $f\rightarrow 0$. In this plot $\rho_{\mathrm i}/\rho_{\mathrm e}=3$. }
\label{LdolambdaoR} \end{figure}

Next we study, using the analytical results given in this Section, how the damping
per wavelength depends on the dimensionless frequency of the driver $f$, defined by
Eq.~(\ref{domega}), for particular cases (see Figure~\ref{Ldolambda}). We see that
the wider the layer the more efficient the attenuation (smaller damping per
wavelengths), in agreement with the analytical results in the TT approximation. For
small $f$ the damping per wavelength tends to the TT value (see dotted line). The
value of $L_{\mathrm D}/\lambda$ increases monotonically with $\omega$. The
deviation with respect to the TT results is smaller for thick layers. In the TT
approximation, which for propagating waves, is the low frequency approximation,
$L_{\mathrm D}/\lambda$ is independent of frequency, since this approximation does
not take into account the variation of frequency, i.e., the frequency is only
presumed to be small. The analytical results for $L_{\mathrm D}/\lambda$ beyond the
TT approximation, take into account the dependence on frequency. The value of
$L_{\mathrm D}/\lambda$ now undergoes a moderate increase when we move from low to
high frequencies. However, the really interesting quantity to calculate is the
damping length itself. In order to make the dependence of $L_{\mathrm D}$ on
frequency more explicit we follow the same line of reasoning in
Section~\ref{TTnonunif} and rewrite Eq.~(\ref{damplength}) as
\begin{equation}
\frac{L_{\mathrm D}}{\lambda}=\xi_{EW},
\end{equation}
where the quantity $\xi_{EW}$ now depends on the parameters of the equilibrium
model and the characteristics of the wave itself. Again $\lambda=
2\pi/k_z$ and $k_z$ is related to the frequency by the dispersion relation
Eq.~(\ref{noTTdisper}). Since we have abandoned the TT approximation, a simple
analytical formula that relates $k_z$ to $\omega$ is not readily available but we
can always write
\begin{equation}
k_z R= f \, \psi (f),
\end{equation}
where $f$ is the dimensionless frequency defined in Eq.~(\ref{domega}) and
$\psi(f)$ is a function that we can determine numerically. Hence, implementing
the function $\psi(f)$, we have that
\begin{equation}
\frac{L_{\mathrm D}}{R}=2 \pi\frac{\xi_{EW}}{\psi(f)}\, \frac{1}{f},
\label{dlEW}
\end{equation}
which in the low frequency limit is equivalent to the TGV relation given by Eq.~(\ref{ldoRtt} ) for $m=1$, i.e.,
\begin{equation}
\xi_{EW}\rightarrow \xi_E \quad \mathrm{and} \quad \psi \rightarrow \sqrt{\frac{\rho_{\mathrm i}+\rho_{\mathrm e}}{2\rho_{\mathrm i}}}
\end{equation}
as $f\rightarrow 0$, where $\xi_E$ is defined by Eq.~(\ref{xidef}). In
Eq.~(\ref{dlEW}), governing the damping length for propagating waves, since
$\psi(f)$ and $\xi_{EW}$ are slowly varying functions of $f$, the main
dependence of $L_{\mathrm D}$ on $f$ is contained in the factor $1/f$, as in the
low frequency limit shown by Eq.~(\ref{ldoRtt}). This is illustrated in
Figure~\ref{LdolambdaoR}, where $L_{\mathrm D}/R$ is plotted as a function of
$f$ for several values of $l/R$. The dependence of the curves on $1/f$ is
very clear and the non-TT and TT solutions (compare the solid with the dotted lines)
tend to overlap in the limit of $f\rightarrow 0$, where both cases are accurately described by the TGV relation. Note that even for
$f\rightarrow 1$ the TGV relation still describes quite well the behaviour of the
damping length with frequency. Again, the potential of resonant absorption as a
frequency filter is clearly demonstrated in Figure~\ref{LdolambdaoR}, with high frequency waves being damped in
shorter spatial scales than low frequency waves (for fixed $l/R$).

\section{Resistive calculations}
\label{RES}

The results based on TB approximation, described in the previous
Sections, are compared with the full resistive calculations \citep[see][for the
details about the method]{terretal06b}. This comparison is useful since in the
resistive eigenvalue problem there are no implicit assumptions about the TT or TB approximation, i.e., the TT and TB approximation are not
used. The resistive calculation applies to any frequency whether it is small
compared to $v_{\mathrm {Ai}}/R$ or not and also to equilibrium models that have
a thin non-uniform layer or are fully non-uniform. We have solved the
eigenvalue problem (for $\omega$) and have used Eq.~(\ref{ldtdgr}) to translate
from temporal damping to spatial damping. Due to the inclusion of
resistivity, the eigenvalue problem for the wavenumber is more difficult to
solve than the eigenvalue problem for the frequency. The results of the
resistive calculations are shown in Figures~\ref{Ldolambda}
and~\ref{LdolambdaoR} where the damping per wavelength and the damping length
are plotted (see dashed lines) as a function of the frequency of the driver. The
agreement between the resistive computations and the analytical or
semi-analytical methods is very good. The small differences are due to the fact
that in the analytical approximations we have assumed that the resonance is
always located at $r=R$ and we have calculated the derivative of the density at
this position. This explains the small deviations from the resistive
estimations. A more precise determination could be done calculating the exact
location of the resonance  and then using a slightly modified version of
Eq.~(\ref{noTTdisperres}), but  since the analytical approximations that we have
already derived are quite satisfactory there is no pressing need to explore this
issue further.

\section{Conclusions and discussion}

The spatial damping due to resonant absorption of driven kink waves has been
investigated. The main conclusion of the work is that the damping length of propagating kink waves due resonant absorption is a monotonically
decreasing function of frequency. The TGV relation for kink waves was derived, demonstrating that for low frequencies the damping length is exactly inversely proportional to frequency. In the high frequency range the TGV relation continues to be an excellent approximation of the actual dependency of the damping length on frequency. Certainly, for all physically relevant frequencies the dependency of damping length on frequency is accurately described by
the TGV relation. This dependency means that resonant absorption is selective as it favours low frequency waves and can efficiently remove high frequency waves from a broad band spectrum of kink waves. This has great significance for solar atmospheric kink waves, since high frequency waves will tend to lose more power than their low frequency counterparts before reaching high altitudes in the solar corona, with the exact percentage power loss depending on the properties of the equilibrium, in particular the width of the non-uniform layer and steepness in variation of the local Alfv\'{e}n speed. With respect to mode conversion, the process of resonant absorption will cause the higher frequency waves to be attenuated more because the global kink mode will be converted into
localised Alfv\'enic modes at lower heights. If the energy of these Alfv\'enic
motions is eventually dissipated, then resonant absorption should  produce a
characteristic distribution of the energy as a function of height in the solar
atmosphere. This could have important consequences with respect to the spatial distribution of wave heating in the solar atmosphere.

It was also shown that spatial and temporal damping are basically equivalent. In the TT approximation the damping per period and the damping per wavelength are exactly the same. The differences in
these two quantities arise in the regime where the TT is not valid, but even in
this situation it is easy to relate the spatial and the temporal damping rates
through the group and phase speeds of the kink MHD waves. This allows us to
translate the results from the temporally damped waves ($\omega$ complex, $k$
real) to spatially attenuated waves ($\omega$ real, $k$ complex) due to
resonant absorption. This mechanism requires the frequency of the driver to be
between the internal and the external Alfv\'en frequency of the tube. This
might seem a very restrictive condition, but in fact it is just the opposite. In
the driven problem the frequency is fixed but the system chooses the proper
wavelength (along the waveguide) to accommodate the kink mode in the tube. This
kink mode generated at the base of the loop propagates upwards along the tube
and at the same time is attenuated due to the inhomogeneity at the tube
boundary.

An interesting result is that both the damping length in the spatial problem and
the damping time in the temporal problem are always smaller than in the TT
approximation (when $f\rightarrow 0$), meaning that waves with short wavelengths or
large frequencies are always more efficiently damped. Note that the observations of
standing kink waves observed with TRACE and for the propagating kink waves detected
with the CoMP instrument are precisely in the regime where the TT is applicable,
i.e., where the waves are less affected by resonant absorption. The fact that
overtones of standing kink waves and high frequency propagating waves have proved
difficult to detect may be a direct consequence of the filtering due to resonant
absorption. From a different perspective, we have also shown that the damping per
wavelength (and the damping per period) has a weak dependence on the frequency.

It is necessary to remark that our results are based on a simple magnetic flux tube
model, i.e., a straight cylinder, with no gravity and pressure. The only
inhomogeneity in this model is plasma density in the radial direction, however this
a characteristic property of solar waveguides that is observed at all atmospheric
heights, e.g., chromospheric magnetic bright points or coronal loops. Thus the
theory of resonant damping of propagating kink waves due to radial plasma density
inhomogeneity offers a natural explanation for the dissipation observed by e.g.,
\citet{tomcmac09}. However, a detailed comparison between the observations and the
expected frequency dependent response due to resonant absorption is needed to
quantify the precise spatial distribution of wave heating due to this mechanism in
the solar plasma, e.g., kink wave dissipation as a function of both frequency and
height. This problem will be the subject of a future work (Verth et al. 2010, in
preparation).

\begin{acknowledgements} J.T. acknowledges the Universitat de les Illes Balears
for a postdoctoral position and the funding provided under projects
AYA2006-07637 (Spanish Ministerio de Educaci\'on y Ciencia). M.G. and G.V.
acknowledge support from K.U.Leuven via GOA/2009-009. \end{acknowledgements}



\end{document}